\documentclass[reprint,amsmath,amssymb,amsbsy,pra
]{revtex4-1}

\usepackage{graphicx}
\usepackage{braket}
\usepackage{bm}
\usepackage{color}
\usepackage{ulem}
\usepackage[colorlinks=true, citecolor=magenta, linkcolor=blue ]{hyperref}

\begin{document}

\newcommand{\txtr}[1]{\textcolor{red}{#1}}
\newcommand{\txtrout}[1]{\textcolor{red}{\sout{#1}}}
\newcommand{\txtb}[1]{\textcolor{blue}{#1}}
\newcommand{\txtg}[1]{\textcolor{green}{#1}}
\newcommand{\txtm}[1]{\textcolor{magenta}{#1}}

\title{Sequential quantum phase transitions in $J_1$-$J_2$ Heisenberg chains with integer spins ($S > 1$):
Quantized Berry phase and valence-bond solids}

\author{Shota Fubasami}
\affiliation{Department of Physics, University of Tsukuba, 1-1-1 Tennoudai, Tsukuba, Ibaraki 305-8571, Japan}

\author{Tomonari Mizoguchi}
\affiliation{Department of Physics, University of Tsukuba, 1-1-1 Tennoudai, Tsukuba, Ibaraki 305-8571, Japan}
\email{mizoguchi@rhodia.ph.tsukuba.ac.jp}

\author{Yasuhiro Hatsugai}
\affiliation{Department of Physics, University of Tsukuba, 1-1-1 Tennoudai, Tsukuba, Ibaraki 305-8571, Japan}

\begin{abstract}
On the basis of a Berry-phase analysis, we study the ground state of the $J_1$-$J_2$ Heisenberg chain for $S=2,3,4$. 
We find that changes of the Berry phase occur $S$ times for spin-$S$ systems, indicating the sequential phase transitions. 
The sequential phase transitions are associated with the change of valence-bond configurations of the ground states.
To demonstrate this numerically, 
we connect the $J_1$-$J_2$ Heisenberg Hamiltonian with the generalized Affleck-Kennedy-Lieb-Tasaki model,
and show that the different phases are connected to the different valence-bond-solid states. 
\end{abstract}
\maketitle
\section{Introduction}
Over the past few decades, quantum spin chains 
with integer spin quantum numbers have provided a fertile ground for 
seeking topological phenomena in many-body physics. 
The $S=1$ Heisenberg model with the nearest-neighbor (NN) interaction is a representative example. 
Its ground state is unique, has a short-range spin correlation, and hosts a finite energy gap on top of it. 
This state, which is called the Haldane state~\cite{Haldane1983,Affleck1989}, is nowadays regarded as a representative example 
of the symmetry-protected topological phase~\cite{Pollmann2010,Pollmann2012,Senthil2015,Wen2017} in many-body quantum systems. 
Clearly, this phase is not characterized by the Ginzburg-Landau-type local order parameters.
Indeed, for the $S=1$ Heisenberg model, the ground state is characterized by the hidden valence-bond solid (VBS) state
as inferred from the related model introduced by Affleck, Kennedy, Lieb, and Tasaki (AKLT)
whose exact ground state is the VBS state~\cite{Affleck1987,Affleck1988}.  
To capture such hidden structures in symmetry-protected topological phases, one needs to employ either non-local order parameters~\cite{Affleck1988, denNijs1989,Kennedy1992}
or so-called topological order parameters~\cite{Thouless1982,Kohmoto1985,Wen1989,Nakamura2002,Kane2005,Hatsugai2006,Fu2007,Tasaki2018}.   
Also, in many cases, 
characteristic boundary states for the open systems, 
e.g. the free end spins with $S=1/2$ for the $S=1$ Heisenberg model~\cite{Kennedy1990},
are the hallmark of the symmetry-protected topological phase~\cite{Halperin1982,Hatsugai1993}. 

The richness of the phases becomes even more abundant 
when considering the models beyond the simple NN-Heisenberg models,
such as introducing the bond alternation~\cite{Affleck1987_bond,Guo1990,Kato1994} and a biquadratic term~\cite{Affleck1987,Affleck1988,Yip2003,Lauchli2006}.
In the present work, we focus on the $J_1$-$J_2$ Heisenberg model, 
where the next-nearest-neighbor (NNN) interaction ($J_2$) is introduced in addition to the NN interaction ($J_1$),
resulting in the frustration. 
In the literatures~\cite{Pati1996,Kolezhuk1996,Hikihara2000,Kolezhuk2002,Pixley2014,Chepiga2016,Chepiga2016_2,Chepiga2017}, 
it was found that, upon changing 
the ratio of $J_2$ to $J_1$, 
the first-order quantum phase transition occurs near $J_2/J_1 \sim 0.75$. 
Indeed, the competition between $J_1$ and $J_2$ plays an important role in this phase transition,
namely, it is thought to be associated with the change in the corresponding VBS states~\cite{Kolezhuk2002}.
To be specific, the valence bonds live in the NN bonds for weak $J_2$, whereas they live in the NNN bonds for strong $J_2$.
Recently, Chepiga \textit{et al.} found
the change in the $\mathbb{Z}_2$ Berry phase~\cite{Berry1984, Hatsugai2006} across the phase transition,
which is compatible with the change in VBS patterns~\cite{Chepiga2016_2}.
 
Motivated by these works, in the present paper, 
we investigate the ground state of the $J_1$-$J_2$ Heisenberg chain for higher-order integer spins, $S=2,3,4$. 
The models (up to $S=2$, including $XY$ and $XXZ$ models) were investigated in 
Refs.~\onlinecite{Roth1998,Lecheminant2001,Hikihara2001}, and 
Ref.~\onlinecite{Roth1998} indicates the existence of the phase transition 
for the $S=2$ Heisenberg chain, through analyses of the end spins for open systems. 
Based on these results, the goal of our study is to characterize the phase transitions by means of Berry-phase analysis. 
We have found sequential changes in the Berry phase, 
which occur $S$ times for the spin-$S$ $J_1$-$J_2$ model, indicating sequential topological phase transitions.
We further associate these transitions with the change in the VBS pictures of the ground states. 
It was argued in Ref.~\onlinecite{Kolezhuk2002} that the phase for $J_2 < J_2^c$ and $J_2 > J_2^c$
correspond to the NN-VBS state and NNN-VBS state [see Fig.~\ref{Fig4}(a) and \ref{Fig4}(b)], respectively, for $S=1$.
To verify this scenario and further generalize it to higher spins, 
we consider models in which the Heisenberg model is continuously connected to 
the generalized AKLT models with various connectivity of singlets, and see whether the ground states are adiabatically connected upon changing the parameter.
We find that the different phases in the $J_1$-$J_2$ 
Heisenberg chain are connected to the different VBS states,
and that the number of phases for higher $S$ values coincides with the number of possible VBS patterns. 
\begin{figure}[b]
\begin{center}
\includegraphics[width= 0.95\linewidth]{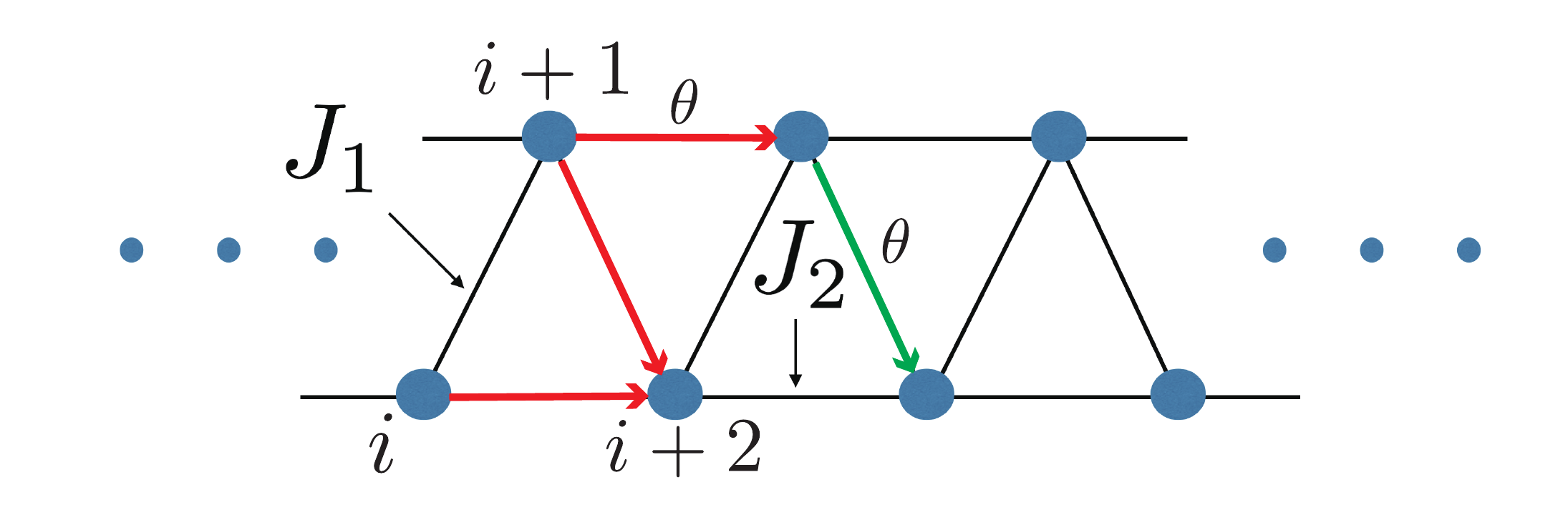}
\caption{The $J_1$-$J_2$ Heisenberg chain. 
Red and green lines represent bonds where the twist of the interaction is introduced
for the results in Secs.~\ref{sec:result1} and \ref{sec:result2}, respectively. }
\label{Fig1}
\end{center}
\end{figure}

The rest of this paper is organized as follows.
In Sec.~\ref{sec:model}, we first introduce 
the $J_1$-$J_2$ antiferromagnetic Heisenberg chain, which is the main focus of this paper. 
We also describe our method, namely, detection of the topological phase transitions by using the $\mathbb{Z}_2$ Berry phase. 
The details of the numerics are also elucidated. 
The main results of this paper are reported in Sec.~\ref{sec:result}.
First, we present the results of the $\mathbb{Z}_2$ Berry phase for $S=2,3,4$.
We show that there appear sequential changes in the $\mathbb{Z}_2$ Berry phase as a function of $J_2/J_1$.
We then move on to the analysis of the Heisenberg chain and generalized ALKT models with various connectivity of singlets.
Finally, we present a summary of this paper in Sec.~\ref{sec:summary}. 

\section{Model and method \label{sec:model}}
\subsection{$J_1$-$J_2$ Heisenberg chain}
We consider the $J_1$-$J_2$ antiferromagnetic Heisenberg chain for $S = 1, 2, 3$ and $4$.
The Hamiltonian reads
\begin{equation}
H^{(S)}_{\mathrm{H}} = \sum_{i = 1}^{N} J_1 \bm{S}_i \cdot \bm{S}_{i+1} + J_2 \bm{S}_i \cdot \bm{S}_{i+2},
\end{equation}
where $\bm{S}_i$ is a spin operator at site $i$, and $N$ is the number of sites.
The exchange parameters, $J_1$ and $J_2$, are set to be non-negative.  
Hereafter, unless otherwise noted, we set $J_1 = 1$, as a unit of energy. 
We impose a periodic boundary condition, $ \bm{S}_{i+N} =  \bm{S}_{i}$.
See Fig.~\ref{Fig1} for a schematic of the model. 
We note that, throughout this paper, 
we consider finite chains
whose ground states are accessible by exact diagonalization (see Sec.~\ref{sec:numerics} for details). 

\subsection{$\mathbb{Z}_2$ Berry phase}
For gapped quantum systems, the Berry phase~\cite{Berry1984} with respect to the twist angles of boundary conditions 
is enforced to be quantized when the Hamiltonian has some discrete symmetries.
Indeed, it has served as a powerful tool to capture the topological phases in various systems of finite system size~\cite{Hatsugai2006, Hatsugai2007,Maruyama2007,Hirano2008,Hirano2008_2,Maruyama2009,Hatsugai2011,Chepiga2013,Motoyama2013,Kariyado2014,Kariyado2015,Chepiga2016_2,Kariyado2018,Motoyama2018,Maruyama2018,Palumbo2019,Kawarabayashi2019}. 

For quantum spin systems having time-reversal symmetry, 
$\mathbb{Z}_2$ Berry phase serves as a topological order parameter~\cite{Hatsugai2006}.
The Berry phase is defined with respect to the twist angle, $\theta$,
which modulates the Hamiltonian in the following manner. 
The twist at the bond $\langle i,j \rangle$ 
with the angle $\theta \in [0,2\pi]$ is introduced 
by replacing 
$ \bm{S}_i \cdot  \bm{S}_j$ in the Hamiltonian with
$\frac{1}{2} \left( e^{i\theta} S_i^+ S_j^- + e^{- i\theta} S_i^- S_j^+ \right) + S_i^z S^z_j$.   
We label the Hamiltonian with the twist $H(\theta)$.
Suppose that the ground state of $H(\theta)$, $\ket{\Phi_0(\theta)}$, remains unique upon varying $\theta$.
Then, one can define the Berry connection,
\begin{eqnarray}
A (\theta) = \bra{\Phi_0(\theta)} \partial_{\theta} \ket{\Phi_0(\theta)}, 
\end{eqnarray}
and the corresponding Berry phase
\begin{eqnarray}
i\gamma =\int_{0}^{2\pi} d\theta \hspace{.5mm} A (\theta). \label{eq:berry}
\end{eqnarray}
Note that $\gamma$ is quantized as $\gamma = 0, \pi$ (mod $2\pi$) due to the time-reversal symmetry. 
Namely, the twisted Hamiltonian satisfies $\mathcal{T} H(\theta) \mathcal{T}^{-1} = H(- \theta)$, which leads to $\gamma = -\gamma$ (mod $2\pi$).
Roughly speaking, $\gamma = \pi \hspace{.5mm} (0)$ indicates that there exists an odd (even) number of hidden spin singles in the twisted bond(s)~\cite{Hatsugai2006,Hirano2008,Chepiga2016_2}. 
Therefore, it is useful for seeking the corresponding VBS picture of the ground state.

\subsection{Numerical calculation of $\mathbb{Z}_2$ Berry phase \label{sec:numerics}}
We numerically calculate the ground state $\ket{\Phi_0(\theta)}$ 
by exact diagonalization.  
To obtain $\gamma$, the integration in Eq. (\ref{eq:berry}) is approximated by the summation, as
\begin{eqnarray}
i \gamma \sim  \sum_{n=0}^{N_{\mathrm{m}}-1} \Delta \theta \langle \Phi_0(\theta_n)  \ket{  \partial_{\theta} \Phi_0(\theta_{n})},
\end{eqnarray}
where $\Delta \theta = \frac{2 \pi}{N_{\mathrm{m}}}$ and $\theta_n= n\Delta \theta$; 
$N_{\mathrm{m}}$ is the number of meshes in a space of $\theta$. 
To avoid the gauge-fixing problem, the summation can be further approximated as follows~\cite{Fukui2005,Hatsugai2006}.
First, we define a quantity $U_n = \langle \Phi_0(\theta_n) | \Phi_0(\theta_{n+1}) \rangle$,
which can be approximated up to $O(\Delta \theta)$ as 
$U_n = 1 + \Delta \theta  \langle \Phi_0(\theta_n) | \partial_{\theta} \Phi_0(\theta_{n}) \rangle +O(\Delta \theta^2 )$.
Next, we introduce a gauge-invariant quantity $\log \left[ \prod_{n=0}^{ N_{\mathrm{m}}-1 }U_n   \right]$,
which can be approximated as $\log \left[ \prod_{n=0}^{N_{\mathrm{m}} -1 }U_n   \right] 
= \sum_{n=0}^{N_{\mathrm{m}} -1 } \Delta \theta
\langle \Phi_0(\theta_n) | \partial_{\theta} \Phi_0(\theta_{n}) \rangle  + O(\Delta \theta^2)$,
and thus we obtain $i \gamma \sim \log \left[ \prod_{n=0}^{N_{\mathrm{m}}-1}U_n \right]$.
We have confirmed that the numerical error due to this approximation is at most $\mathcal{O}(10^{-6})$ in units of $2\pi$.

As for the choice of the twisted bonds,
we use a three-bond twist, represented by red arrows in Fig.~\ref{Fig1} for the results in Sec.~\ref{sec:result1}.
This choice is the same as that in the previous work~\cite{Chepiga2016_2}, 
which turned out to reduce the finite-size effect (see Fig.~\ref{Fig:FS}). 
On the other hand, for the results in Sec.~\ref{sec:result2}, 
we employ the single-bond twist, represented by the green arrow in Fig.~\ref{Fig1},
for simplicity of calculations. 
It should be noted that both of the choices of the twisted bonds contain an odd number of NN bonds and an even number of NNN bonds,
so the corresponding Berry phases should be the same, except for deviations due to the finite-size effect.

\section{Results \label{sec:result}}
\subsection{Sequential change of $\mathbb{Z}_2$ Berry phase for $S > 1$ \label{sec:result1}}
\begin{figure*}[t]
\begin{center}
\includegraphics[width= 0.95\linewidth]{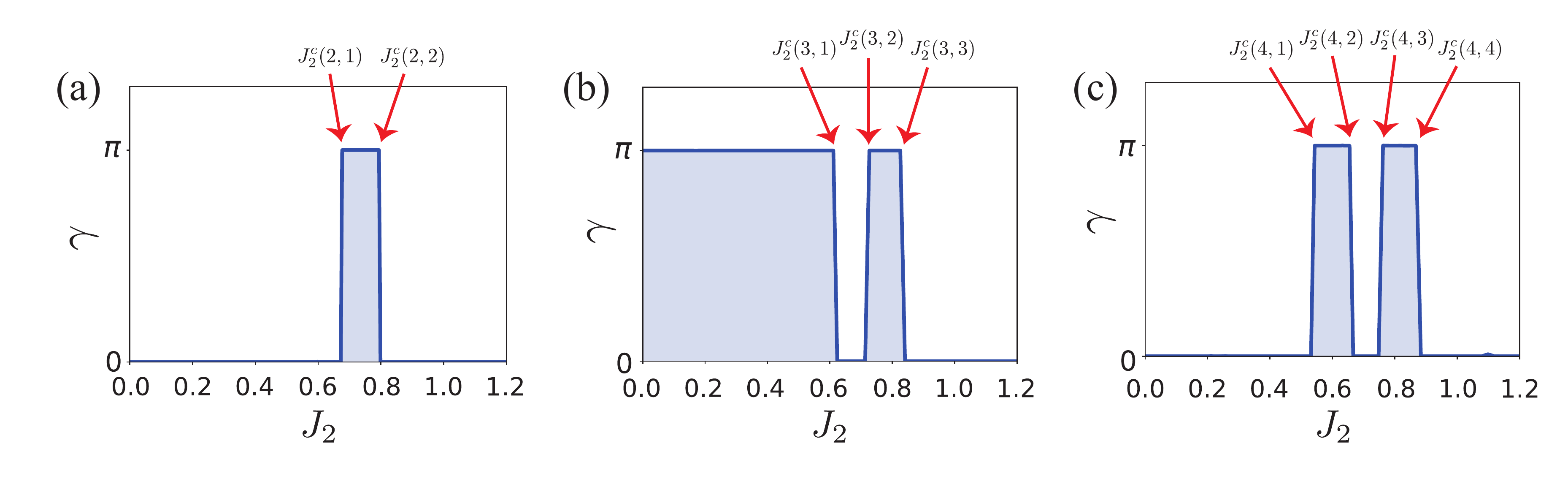}
\caption{$\mathbb{Z}_2$ Berry phase
as a function of $J_2$ for (a) $S=2, N = 10$, (b) $S=3, N=10$
and (c) $S=4, N=8$.
The critical points, $J_2^c(S,n)$, are
$J_2^c (2,1) = 0.6723$, 
$J_2^c (2,2) = 0.7939$ for $S=2$;
$J_2^c (3,1) =0.6112 $, 
$J_2^c (3,2) = 0.7130$,
$J_2^c (3,3) = 0.8257$ for $S=3$;
$J_2^c (4,1) = 0.5316$, 
$J_2^c (4,2) = 0.6552$,
$J_2^c (4,3) = 0.7483$ ,
$J_2^c (4,4) = 0.8667$ for $S=4$.} 
\label{Fig:berry}
\end{center}
\end{figure*}
\begin{figure}[b]
\begin{center}
\includegraphics[width= 0.95\linewidth]{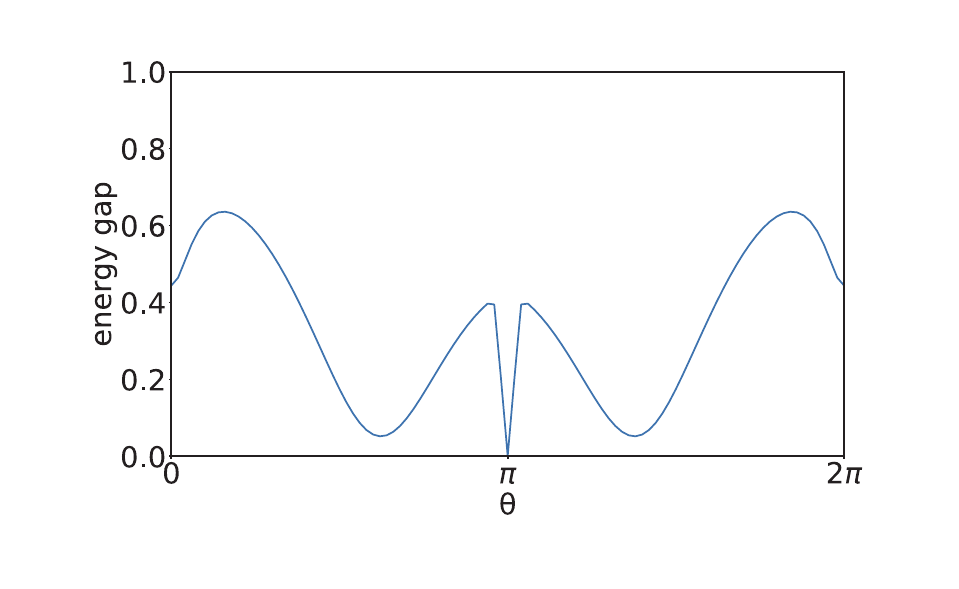}
\caption{
Energy gap as a function of $\theta$ at $(J_1, J_2) = (0.8299, 0.5579)$,
i.e. $J_2 /J_1 = 0.6723$,
 for $S=2$.
}
\label{Fig:gap}
\end{center}
\end{figure}
In this subsection, we discuss the $J_2$ dependence of the $\mathbb{Z}_2$ Berry phase for various values of $S$.
The result for $S=1$ was already uncovered~\cite{Chepiga2016_2}, so we focus on the case with a higher $S$. 

In Fig.~\ref{Fig:berry}, we show the $\mathbb{Z}_2$ Berry phase
as a function of $J_2$ for $S=2,3,4$. 
The system sizes used are presented in the caption. 
We clearly see the changes in $\gamma$
from $0$ to $\pi$ or from $\pi$ to $0$ when changing $J_2$, 
which indicate quantum phase transitions.
Notice that the changes
occur $S$ times for the spin-$S$ model.
For instance, for $S=2$, the Berry phase varies 
as $0 \rightarrow \pi \rightarrow 0$, upon increasing $J_2$.
Henceforth, we label the $n$-th transition point for the spin-$S$ model as $J_2^{c} (S,n)$
(see red arrows in Fig.~\ref{Fig:berry}). 
\begin{figure}[b]
\begin{center}
\includegraphics[width= 0.95\linewidth]{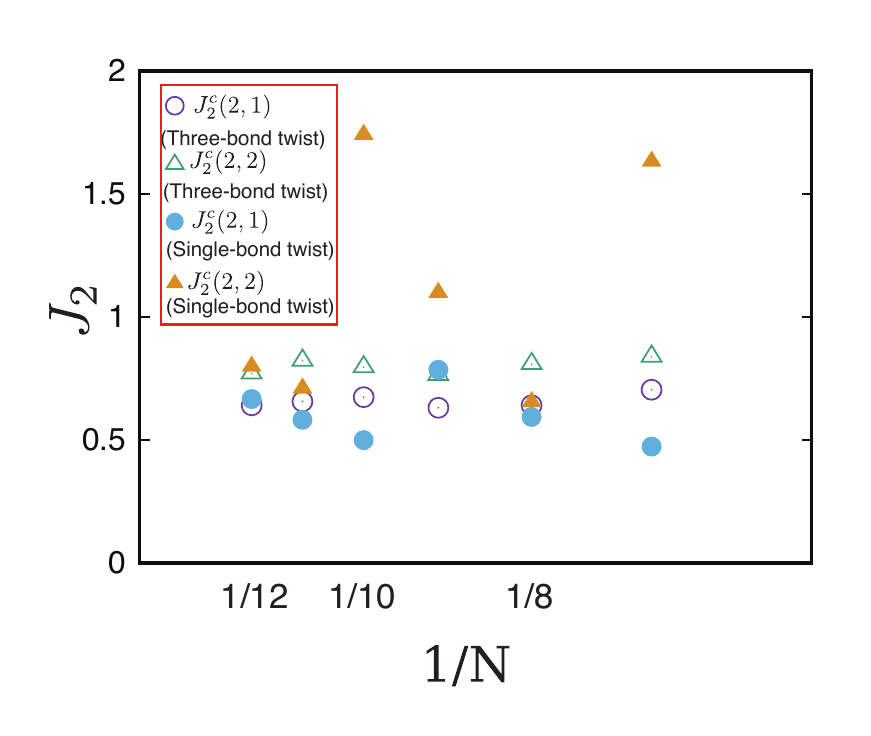}
\caption{Size dependence of the critical points for $S=2$.}
\label{Fig:FS}
\end{center}
\end{figure}

We further find another sign of the phase transition, namely the ``hidden gap closing" when the twist is introduced. 
For the untwisted $J_1$-$J_2$ model, 
the energy gap between the ground state and the first excited state does not close across the change of the Berry phase. 
This behavior was also observed for $S=1$ in the density matrix renormalization group study~\cite{Kolezhuk1996}. 
Our finding is that, at the critical values of $J_2$, the energy gap closes for the twisted model with $\theta = \pi$;
see Fig.~\ref{Fig:gap} for the energy gap as a function of $\theta$ at $J_2^c (2,1)$.
Note that similar behavior is argued in Refs.~\onlinecite{Kitazawa1997} and \onlinecite{Nomura1998} in the context of the field-theoretical analysis.

As we have emphasized, the results shown in Fig.~\ref{Fig:berry} are for a finite size system,
and determination of the critical points in the thermodynamic limit is hampered by the finite size effect,
as pointed out for $S=1$~\cite{Chepiga2016_2}.
As an example, we show the size dependence of $J_2^c$ for $S=2$ in Fig.~\ref{Fig:FS}.
For comparison, we also plot $J_2^c$ obtained from the single-bond twist, which we employ in the next subsection.
Clearly, the convergence of $J_2^c$ is not obtained up to $N=12$.
We also see that the choice of twisted bonds quantitatively affects the $J_2^c$.
Therefore, one needs to employ an alternative method to obtain the precise critical points in the thermodynamic limit.

Although the existence of phase transitions is captured in the $\mathbb{Z}_2$ Berry phase, 
the physical pictures of the phases can not be identified directly, which we elucidate in the next subsection. 

\subsection{Connection to the generalized AKLT models \label{sec:result2}}
\begin{figure}[b]
\begin{center}
\includegraphics[width= 0.95\linewidth]{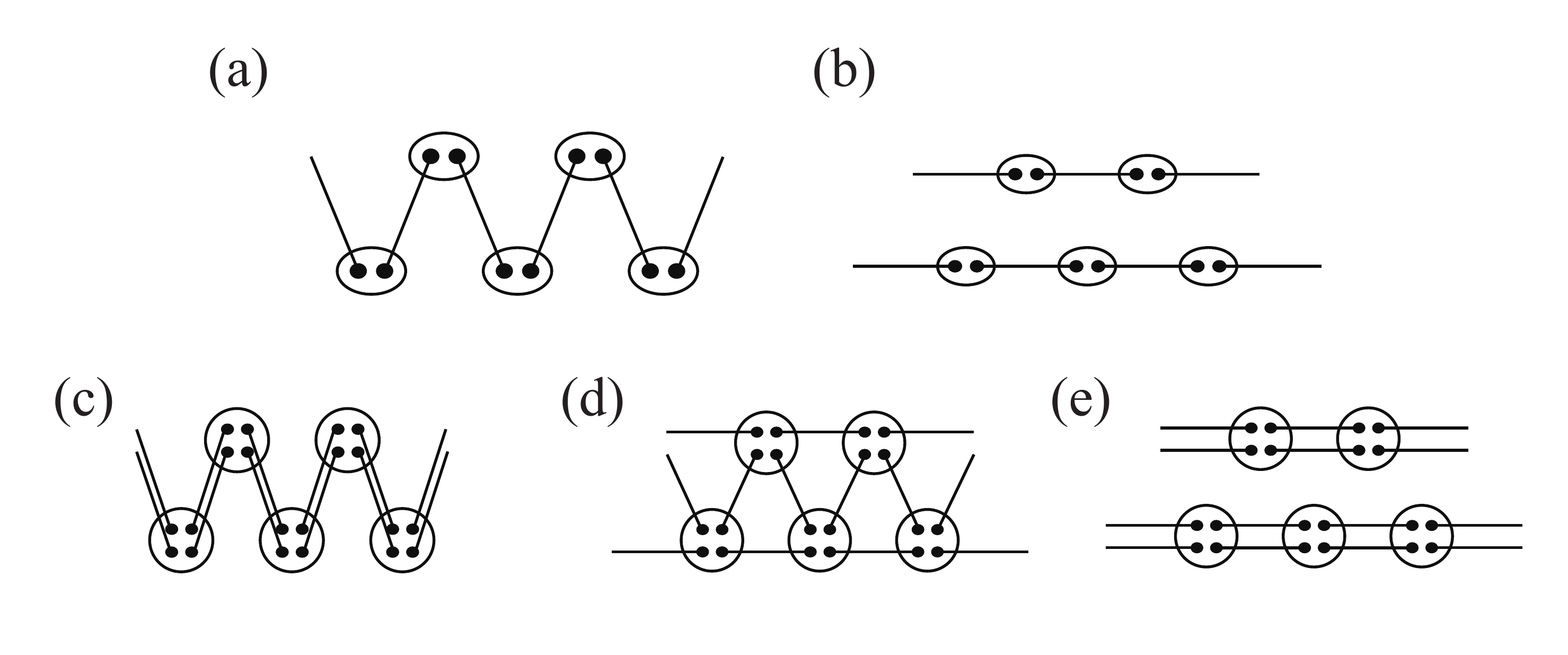}
\caption{Schematics of the VBS ground states of  
(a) $H_{\mathrm{AKLT}}^{(1), \mathrm{(i)}}$,
(b) $H_{\mathrm{AKLT}}^{(1), \mathrm{(ii)}}$,
(c) $H_{\mathrm{AKLT}}^{(2), \mathrm{(i)}}$,
(d) $H_{\mathrm{AKLT}}^{(2), \mathrm{(ii)}}$,
and (e) $H_{\mathrm{AKLT}}^{(2), \mathrm{(iii)}}$.
Small black squares, lines, and open circles
represent 
the spin $1/2$, the spin singlets, 
and the symmetrization, 
respectively. 
}
\label{Fig4}
\end{center}
\end{figure}
To understand the origin of 
the quantum phase transitions, 
we continuously connect the $J_1$-$J_2$ Heisenberg model with
the generalized AKLT models whose ground states are the VBS states, up to $S=2$. 

The possible VBS states for the $J_1$-$J_2$ model are listed in Fig. \ref{Fig4}.
For $S=1$, there are only two possible VBS states.
One has spin singlets in the NN bonds [NN-VBS state; see Fig.\ref{Fig4}(a)]; 
the other in the NNN bonds [NNN-VBS state; see Fig.\ref{Fig4}(b)].
For $S=2$, on the other hand, there is another state, in addition to the NN-VBS and NNN-VBS states, 
in which the singlets live in both NN bonds and NNN bonds [Fig.\ref{Fig4}(d)].
We refer to this state as the intermediate(I)-VBS state. 
\begin{figure}[b]
\begin{center}
\includegraphics[width= 0.98\linewidth]{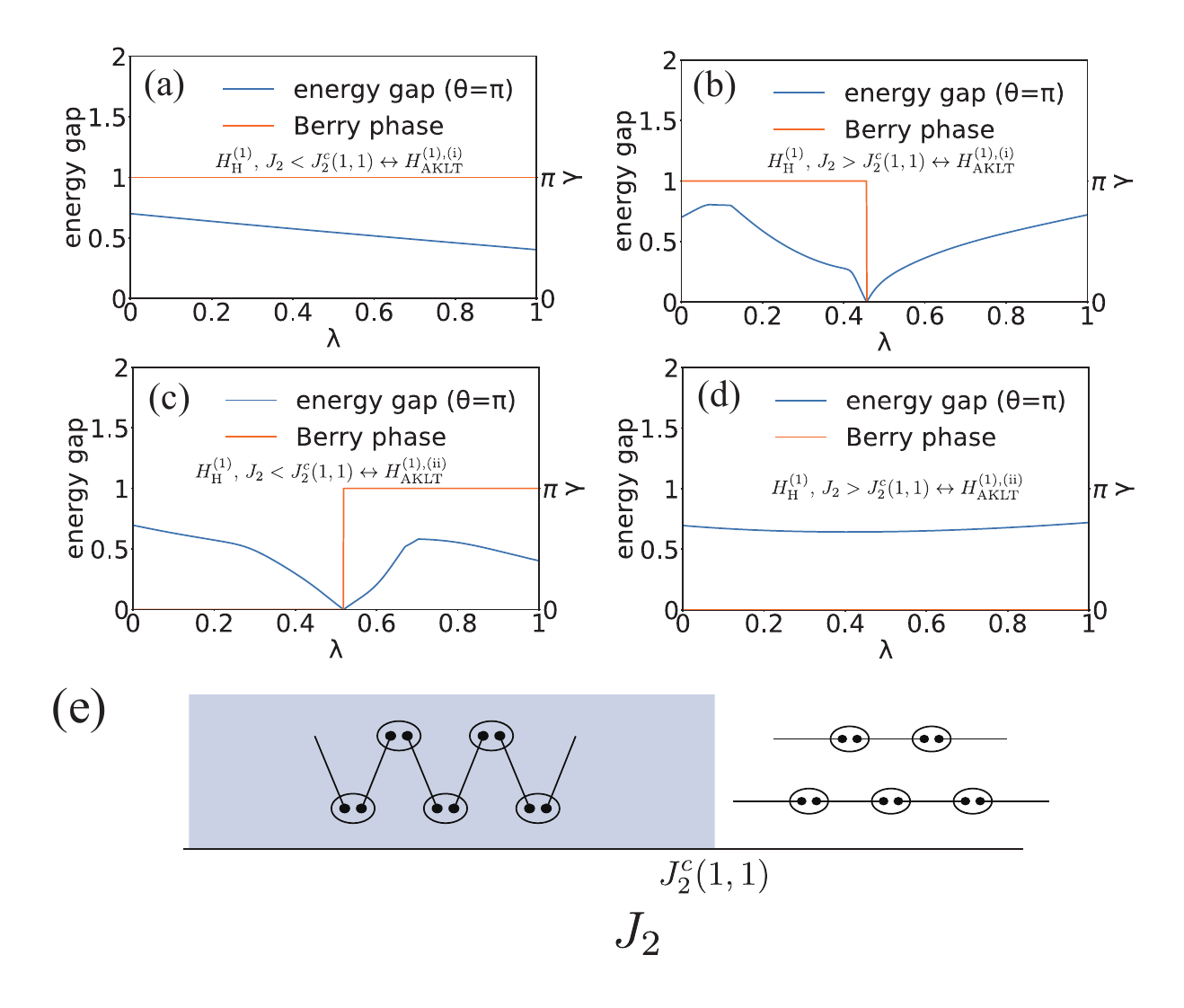}
\caption{(a)-(d) $\lambda$ dependence 
of the energy gap at $\theta = \pi$ and the Berry phase
of the Hamiltonian, Eq. (\ref{eq:twistHam_lambda}) for $S=1$.
Parameters and choices of $\eta$ are (a) $J_1=1, J_2=0$ and $\eta=$ (i);
(b) $J_1=0, J_2=1$ and $\eta=$ (i);
(c) $J_1=1, J_2=0$ and $\eta=$ (ii);
and (d) $J_1=0, J_2=1$ and $\eta=$ (ii). 
The system size is $N = 12$. 
The gap closing occurs at $\lambda = 0.4565$ in (b), and 0.5185 in (c).
(e) Schematic phase diagram of the $J_1$-$J_2$ model for $S=1$ with respect to the VBS picture. Shaded area represents $\gamma = \pi$. }
\label{Fig_S1_VBS}
\end{center}
\end{figure}
To obtain the desirable AKLT models, one can follow the recipe to write down the Hamiltonian with respect to the projection operators that favor the corresponding VBS states~\cite{Arovas1988,Auerbach}. 
In what follows, we employ
\begin{eqnarray}
  H_{\mathrm{AKLT}}^{(1), \mathrm{(i)}} &= & 
2\sum_{i=1}^{N} P^1_{2} (i,i+1)
\nonumber  \\
&=&   \sum_{i=1}^{N} \bigg[
  \bm{S}_i \cdot \bm{S}_{i+1} +\frac{1}{3} \left( \bm{S}_i \cdot \bm{S}_{i+1} \right)^2  + \frac{2}{3}
  \bigg],
\end{eqnarray}
and 
\begin{eqnarray}
  H_{\mathrm{AKLT}}^{(1), \mathrm{(ii)}}
 &= &
2\sum_{i=1}^{N} P^1_{2} (i,i+2)
\nonumber \\  
& = &  \sum_{i=1}^{N}
\bigg[
  \bm{S}_i \cdot \bm{S}_{i+2} +\frac{1}{3} \left( \bm{S}_i \cdot \bm{S}_{i+2} \right)^2  + \frac{2}{3}
  \bigg],
\end{eqnarray}
for $S=1$; similarly, 
\begin{eqnarray}
  H_{\mathrm{AKLT}}^{(2), \mathrm{(i)}} &= &
10\sum_{i=1}^{N}  \left(\frac {1}{7}    P^2_{3} (i,i+1)  +  P^2_{4} (i,i+1) \right) 
  \nonumber  \\
&= &  \sum_{i=1}^{N} 
\bm{S}_i \cdot \bm{S}_{i+1} +\frac{2}{9} \left( \bm{S}_i \cdot \bm{S}_{i+1} \right)^2 \nonumber  \\ 
&+&\frac{1}{63} \left( \bm{S}_i \cdot \bm{S}_{i+1} \right)^3 + \frac{10}{7},
\end{eqnarray}
\begin{eqnarray}
  H_{\mathrm{AKLT}}^{(2), \mathrm{(ii)}}
  &=&
28 \sum_{i=1}^{N} \sum_{\delta = 1,2}   P^2_{4} (i,i+\delta  ) 
\nonumber   \\
  &=&  
\sum_{i=1}^{N} \sum_{\delta = 1,2}
\bm{S}_i\cdot \bm{S}_{i+\delta}
+\frac{7}{10}(\bm{S}_{i}\cdot\bm{S}_{i + \delta })^2 \nonumber \\
&+&\frac{7}{45}(\bm{S}_{i}\cdot\bm{S}_{i + \delta })^3 
+\frac{1}{90}(\bm{S}_{i}\cdot\bm{S}_{i + \delta })^4,
\end{eqnarray}
and
\begin{eqnarray}
  H_{\mathrm{AKLT}}^{(2), \mathrm{(iii)}}
 &= &
10\sum_{i=1}^{N}  \left(\frac {1}{7}    P^2_{3} (i,i+2)  +  P^2_{4} (i,i+2) \right)
  \nonumber  \\
  &=& \sum_{i=1}^{N} 
\bm{S}_i \cdot \bm{S}_{i+2} +\frac{2}{9} \left( \bm{S}_i \cdot \bm{S}_{i+2} \right)^2 \nonumber  \\
&+&\frac{1}{63} \left( \bm{S}_i \cdot \bm{S}_{i+2} \right)^3 + \frac{10}{7},  \nonumber  \\
\end{eqnarray}
for $S=2$~\cite{Katsura2007}, 
where $P^S_J(i,j)$ is a projection operator for the total spin $J$ of the bond $\langle i,j \rangle $.
The  VBS state is a zero energy state of the Hamiltonian since it does not have any projection for the subspace $J>2S-M$ where $M$ is the number of valence bonds~\cite{Arovas1988,Auerbach}.

Then consider the Hamiltonian with the additional parameter $\lambda \in [0,1]$:
\begin{eqnarray}
 H^{(S)}(\lambda,\theta)  = \lambda H^{(S)}_{\mathrm{H}} (\theta) + \left(1- \lambda \right) H^{(S), \eta}_{\mathrm{AKLT}} (\theta). \label{eq:twistHam_lambda}
\end{eqnarray}
Here $\eta =$ (i), (ii) for $S=1$ and $\eta =$ (i), (ii), (iii) for $S=2$. 
Note that the twist of the Hamiltonian is introduced in
both $H^{(S)}_{\mathrm{H}}$ and $H^{(S), \eta}_{\mathrm{AKLT}}$.
The higher-order terms of $\bm{S}_i \cdot \bm{S}_j $ are twisted as 
$\left(\bm{S}_i \cdot \bm{S}_j \right)^{\alpha} \rightarrow\left[  \frac{1}{2}\left(
e^{i \theta} S_i^+ S_j^- + e^{- i \theta} S_i^- S_j^+  
\right) + S_i^z S_j^z \right]^{\alpha} $
with $\alpha$ being a positive integer. 

We aim to see whether the ground states of $H^{(S)}_{\mathrm{H}}$ and $H^{(S),\eta}_{\mathrm{AKLT}}$ 
are connected by introducing the Hamiltonian, Eq. (\ref{eq:twistHam_lambda}). To this end,
we monitor the energy gap upon varying $(\lambda,\theta)$. 
We also compute the $\lambda$ dependence of the $\mathbb{Z}_2$ Berry phase. 

\begin{figure*}[t]
\begin{center}
\includegraphics[width= 0.95\linewidth]{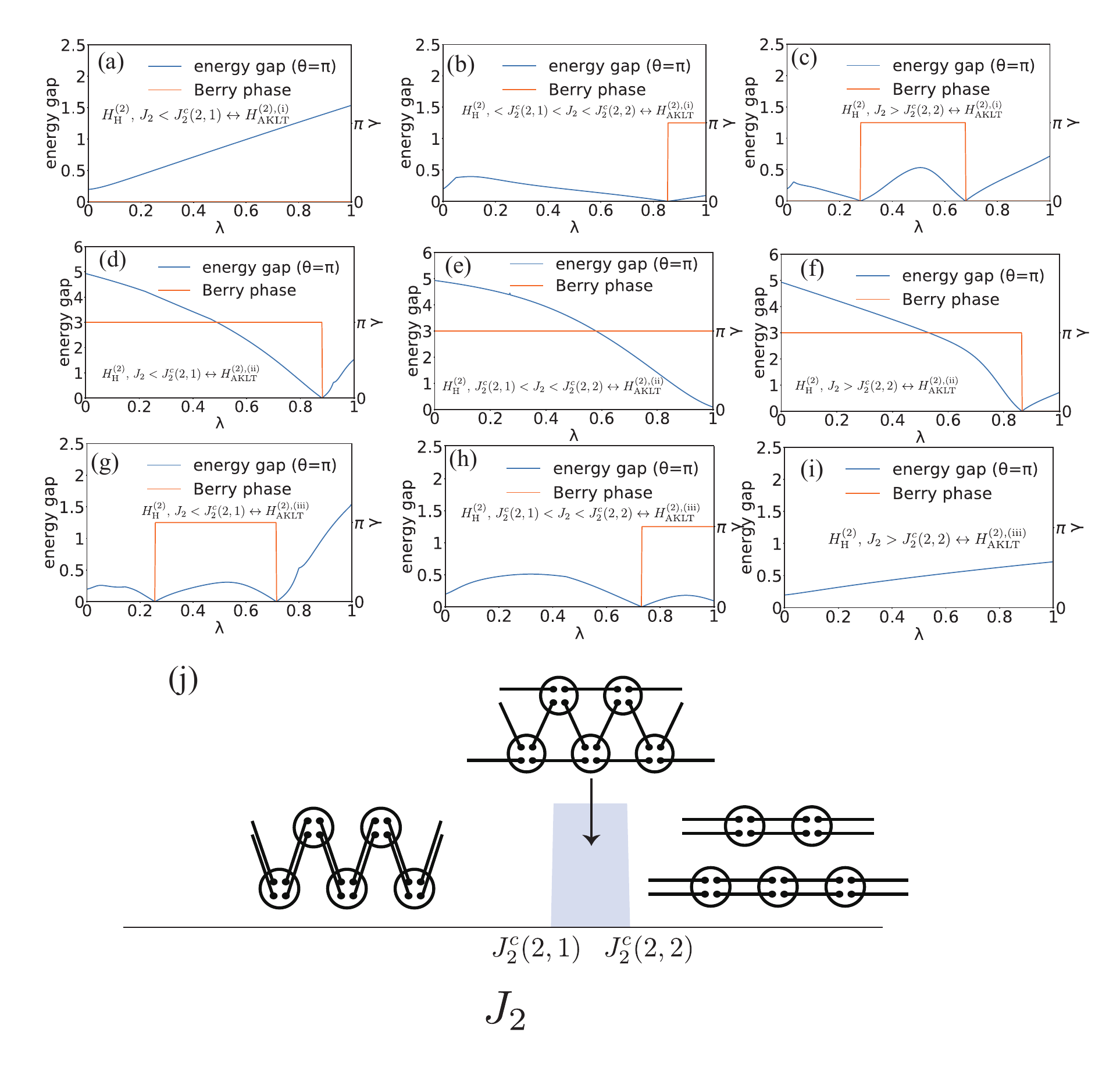}
\caption{
(a)-(i) $\lambda$ dependence 
of the energy gap at $\theta = \pi$ and the Berry phase
of the Hamiltonian, Eq. (\ref{eq:twistHam_lambda}) for $S=2$.
Parameters and choices of $\eta$ are 
(a) $J_1=1, J_2=0$ and $\eta=$ (i);
(b) $J_1=1, J_2=0.942$ and $\eta=$ (i);
(c) $J_1=0, J_2=1$ and $\eta=$ (i);
(d) $J_1=1, J_2=0$ and $\eta=$ (ii); 
(e) $J_1=1, J_2=0.942$ and $\eta=$ (ii);
(f) $J_1=0, J_2=1$ and $\eta=$ (ii);
(g) $J_1=1, J_2=0$ and $\eta=$ (iii);
(h) $J_1=1, J_2=0.942$ and $\eta=$ (iii); 
and 
(i) $J_1=0, J_2=1$ and $\eta=$ (iii). 
The gap closing occurs at 
$\lambda = 0.855$ in (b), and 0.279 and 0.679 in (c),
0.881 in (d), 0.865 in (f), 0.257 and 0.715 in (g), and 0.731 in (h).
The system size is $N = 9$. 
(j) Schematic phase diagram of the $J_1$-$J_2$ model for $S=2$ with respect to the VBS picture.
Shaded area represents $\gamma = \pi$. }
\label{Fig:result_S2_VBS}
\end{center}
\end{figure*}

Let us first look at the results for $S=1$, shown in Fig.~\ref{Fig_S1_VBS}. 
Clearly, the Heisenberg model with $J_2 < J_2^{c} (1,1)$ is connected to $H^{(1), \mathrm{(i)}}_{\mathrm{AKLT}}$ 
without gap closing or a change in the Berry phase [Fig.~\ref{Fig_S1_VBS}(a)],
whereas that with $J_2 > J_2^{c} (1,1)$ is connected to $H^{(1), \mathrm{(ii)}}_{\mathrm{AKLT}}$ [Fig.~\ref{Fig_S1_VBS}(d)]. 
This indicates that the ground state of the Heisenberg model with 
$J_2 < J_2^{c} (1,1)$ [$J_2 > J_2^{c} (1,1)$] 
is in the same phase as the NN-VBS [NNN-VBS] state, which is consistent with the previous study~\cite{Kolezhuk2002}. 
Also, the corresponding VBS states are consistent with the Berry phase~\cite{Chepiga2016_2}, 
because there is an odd (even) number of the singlets in the NN-VBS (NNN-VBS) state on NN bonds.
In contrast, if the Heisenberg model is connected with the incompatible AKLT model,
e.g., the Heisenberg model with $J_2 < J_2^{c} (1,1) $ and $H^{(1), \mathrm{(ii)}}_{\mathrm{AKLT}}$,
a gap closing at $\theta = \pi$ occurs, associated with the change of the Berry phase [Fig.~\ref{Fig_S1_VBS}(b)]. 
These results clearly illustrate that the change in the Berry phase 
in the $J_1$-$J_2$ Heisenberg model upon varying $J_2/J_1$
originates from the phase transition from the NN-VBS state to the NNN-VBS state. 
It is also found that the ground states of $H_{\mathrm{H}}^{(1)}$ with $J_2 < J_2^c(1,1)$
[$J_2 > J_2^c(1,1)$]  is stable against inclusion of $H^{(1), \mathrm{(ii)} }_{\mathrm{AKLT}}$ $ \left(H^{(1), \mathrm{(i)} }_{\mathrm{AKLT}} \right)$
as long as the energy gap at $\theta=\pi$ remains open, 
reflecting the topological stability of these phases. 
This holds even for $S=2$, as we show later.
We remark that the energy gap shows kinklike structures at some points away from the transition point in Figs.~\ref{Fig_S1_VBS}(b) and ~\ref{Fig_S1_VBS}(c). 
These originate from the fact that the level crossing between the first and the second excited states 
occurs at these points; the same is true in Fig.~\ref{Fig:result_S2_VBS}.

Next, we move on to the results for $S=2$, shown in Fig.~\ref{Fig:result_S2_VBS}. 
Again, the different phases for the Heisenberg model are connected to different VBS states:
the NN-VBS state for $J_2 < J_2^{c}  (2,1)$, the I-VBS state for $J_2^{c}  (2,1) < J_2 < J_2^{c}  (2,2)$,
and the NNN-VBS state $J_2 >  J_2^{c}  (2,2)$.
The number of singlets on the NN bonds varies
as $2 \rightarrow 1 \rightarrow 0$ upon increasing $J_2$, 
which is consistent with the profile of the Berry phase, 
$0 \rightarrow \pi \rightarrow 0$,
as shown in Fig.~\ref{Fig:berry}(a).
This result indicates that the sequential phase transitions for $S=2$ 
also originate from the change in the corresponding VBS states. 

Considering the results on $S=1$ and $S=2$,
it is inferred that, for general integer $S$, 
the $S$-time phase transitions for spin-$S$ models coincide with the $S$ patterns of possible VBS states.
Namely, for $J_2 = 0$, $S$ singlets live in the NN bonds, and upon increasing $J_2$, 
the number of singlets in NN bonds decreases as $S \rightarrow S-1, \cdots   \rightarrow0$, and consequently the number of singlets 
in NNN bonds increases as $0  \rightarrow1  \rightarrow \cdots  \rightarrow S$.

\section{Summary \label{sec:summary}}
To summarize, we have investigated the ground state of $J_1$-$J_2$ Heisenberg models with 
higher integer spins, 
by calculating $\mathbb{Z}_2$ Berry phase. 
We reveal that the spin-$S$ model undergoes $S$-time phase transitions. 
We attribute the sequential phase transitions to the change in VBS patterns,
and demonstrate this by analyzing the models in which the $J_1$-$J_2$ Heisenberg model is continuously connected to the generalized AKLT models. 
The resulting VBS pictures are indeed consistent with the $\mathbb{Z}_2$ Berry phase 
with respect to the parity of the number of singlets living in twisted bonds. 

All the results presented in this paper are for finite-size systems 
with small number of sites accessible with exact diagonalization. 
The precise determination of the critical values in the thermodynamic limit or larger-size cases,
which may be different from those for the system size used in this paper,
is outside the scope of the present paper and will be an important future problem. 
The Berry-phase analysis using quantum Monte Carlo simulations,
developed in Refs.~\onlinecite{Motoyama2013} and \onlinecite{Motoyama2018}, will be useful to access 
larger-size systems.

\acknowledgments
Y. H. thanks Kiyohide Nomura for fruitful discussions. 
Part of numerical calculations were carried out on the Supercomputer Center at Institute for Solid State Physics, University of Tokyo. 
This work was supported by JSPS KAKENHI Grants No. JP17H06138 and No. JP16K13845.

\end{document}